\documentclass[11pt,twoside]{article}

\usepackage{asp2006}
\usepackage{epsf}
\usepackage{lscape}

\begin{document}
\title{UV and FUV spectroscopy of the hybrid PG\,1159-type central star NGC\,7094}   

\author{M\@. Ziegler$^1$, T\@. Rauch$^1$, E\@. Reiff$^1$, K\@. Werner$^1$, J.\,W\@. Kruk$^2$, C.\,M\@. Oliveira$^2$}  
\affil{$^1$Institute for Astronomy and Astrophysics, 
           Kepler Center for Astro and Particle Physics, 
           Eberhard Karls University, T\"ubingen, Germany}    
\affil{$^2$Department of Physics and Astronomy, Johns Hopkins University, Baltimore, U.S.A.}    

\begin{abstract} 
Previous studies aiming at the iron-abundance determination in three
PG\,1159 stars (K\,1$-$16, PG\,1159$-$035, NGC\,7094) and a [WC]-PG\,1159 transition
star (Abell\,78) have revealed that no object shows any iron line in the
UV spectrum. The stars are iron-deficient by at least 1\,dex,
typically. A possible explanation is that iron nuclei were transformed
by neutron captures into heavier elements (s-process), however, the
extent of the iron-destruction would be much stronger than predicted by
AGB star models. But if n-captures are the right explanation, then we
should observe an enrichment of trans-iron elements. In this paper we
report on our search for a possible nickel overabundance in one of the
four Fe deficient PG\,1159 stars, namely the central star NGC\,7094. We are
unable to identify any nickel line in HST and FUSE spectra and conclude
that Ni is not overabundant. It is conceivable that iron was transformed
into even heavier elements, but their identification suffers from the
lack of atomic data.
\end{abstract}

\begin{figure}[ht]
\epsfxsize=\textwidth
\epsffile{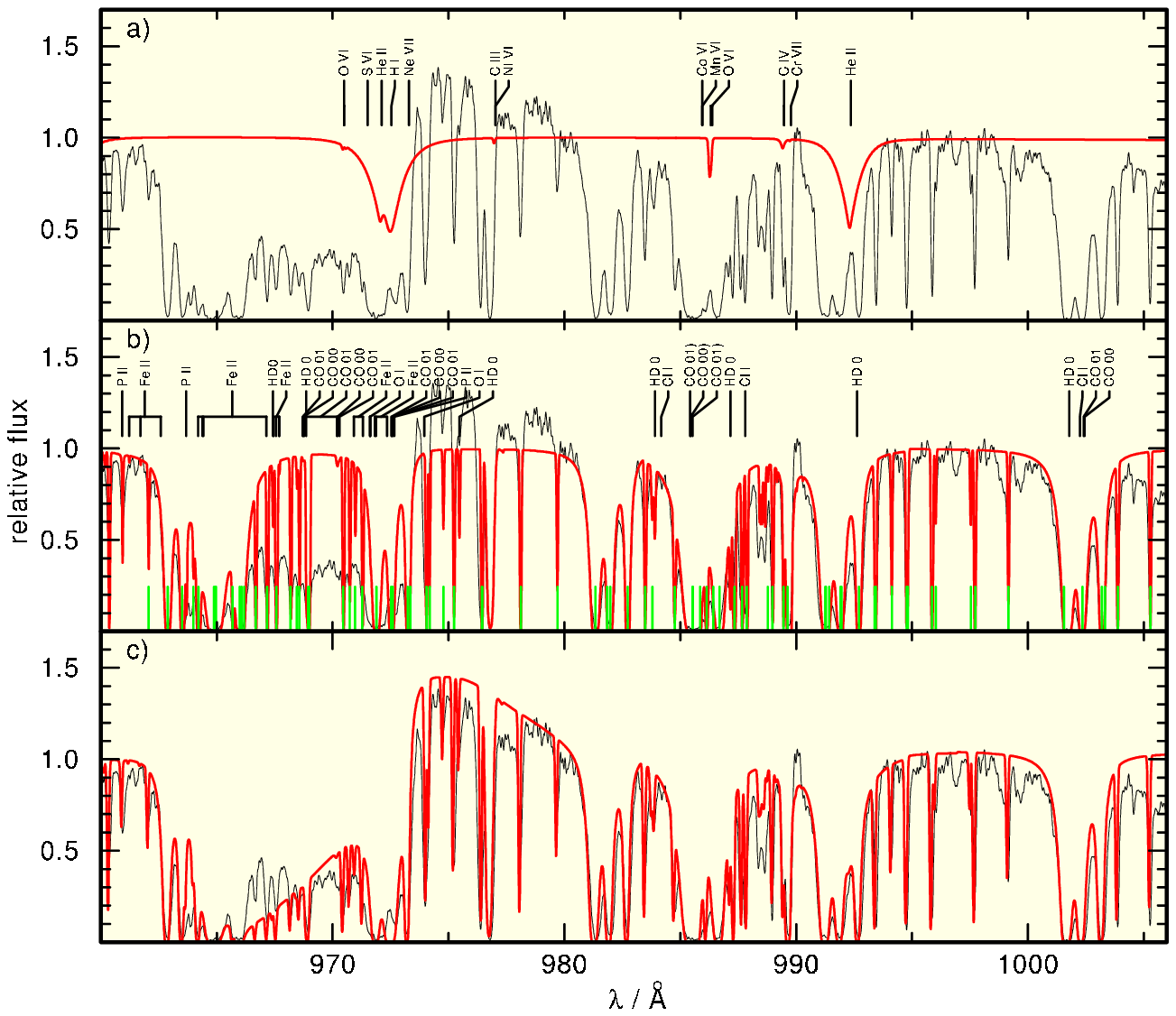}
\caption{Section of the FUSE observation of the CSPN NGC\,7094 compared to our 
a) photospheric (hydrostatic) model, 
b) ISM model, 
c) photospheric (wind) + ISM model. 
Photospheric lines are marked on panel a),
ISM lines are marked on b).
}
\label{fig:ism}
\end{figure}

\section*{Introduction}
PG\,1159 stars form a spectroscopic group of about 40 very hot, H-deficient post-AGB stars. 
FUSE spectroscopy of four PG\,1159 stars, 
namely K\,1-16, NGC\,7094, Abell\,78 ([WC]-PG\,1159), and PG\,1159$-$035 
has shown that these are Fe deficient \citep{mea2002, jea2007}. 
Reasons for this phenomenon may be dust-fractionation (depleted elements are 
incorporated in dust grains and are then expelled from the photosphere by radiation pressure) or the
s-process (n captures during the AGB phase) which transforms Fe into trans-iron elements. 
The detection of s-process signatures in H-deficient post-AGB stars can verify
our current understanding of the synthesis of trans-iron elements in AGB stars: 
these stars contribute to the metal enrichment of the interstellar medium (ISM) 
and hence affect the Galactic chemical evolution.

The Planetary Nebula (PN) NGC\,7094 (PN\,G066.7$-$28.2) has been identified by \citet[K\,1$-$19,][]{k1963}. 
According to \citet{n1999} its distance is 2.2 kpc and 
its apparent size is 102\farcs 5\,$\times$\,99\farcs 4 \citep{tea2003}. 
Its central star (CS) is of so-called hybrid PG\,1159-type, i.e\@. it exhibits H in its spectrum. 
\citet{dwk1997} determined $T_\mathrm{eff}$ = 110\,kK, $\log g = 5.7$, H=0.36, He=0.43, and C=0.21 (mass fractions). 
\citet{mea2002} have shown that the star is Fe deficient. In this analysis, we aim to determine 
the abundances of trans-iron elements in its photosphere. Due to the lack of reliable 
atomic data, we are restricted to Ni, only. 

At the relevant $T_\mathrm{eff}$, Ni\,{\sc vi} and Ni\,{\sc vii} are the dominant ionisation stages in
the line-forming region. Unfortunately, the strongest Ni\,{\sc vii} lines are located at lower wavelengths 
than the FUSE range includes. Most of the Ni\,{\sc vi} lines are accessible in the STIS wavelength range. Based on 
FUSE and STIS observations, we perform a NLTE spectral analysis in order to determine the Ni abundance.

\begin{figure}[ht]
\epsfxsize=\textwidth
\epsffile{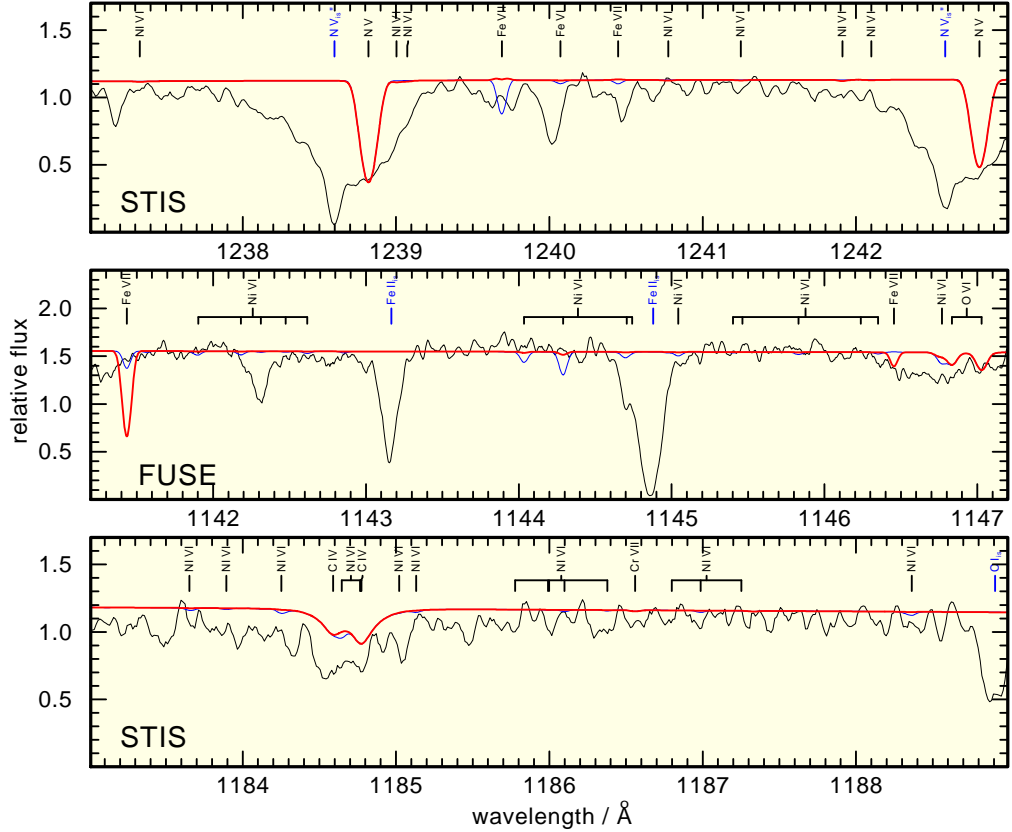}
\caption{
Sections of the STIS and FUSE observations compared with synthetic spectra.
The top panel includes Fe\,{\sc vii} lines which appear slightly too strong (e.g\@. 
Fe\,{\sc vii} $\lambda\,1239.7\,\mathrm{\AA}$) 
in the model at solar abundance (thin, thick $0.1\times$ solar). 
An upper limit for the Fe abundance of $0.025\times$ solar (thin line) can be determined from 
Fe\,{\sc vii} $\lambda\,1141.4\,\mathrm{\AA}$ in the FUSE observation (middle panel, thick line: solar).
Note that no evidence for a Ni overabundance (thick line: solar Ni abundance; thin: $10\times$ solar) can be found 
neither in the FUSE (middle panel, e.g\@. Ni\,{\sc vi} $\lambda\,1144.3\,\mathrm{\AA}$) 
nor in the STIS observation (bottom). 
        }
\label{fig:feni}
\end{figure}

\section*{Spectral Analysis}
The FUV spectrum of NGC\,7094 is strongly contaminated by interstellar absorption 
lines (Fig.~\ref{fig:ism}). These hamper the identification of stellar photospheric lines. 
Thus, modeling the ISM is inevitable for a precise analysis of the FUSE observation. 
Moreover, the FUV and UV spectra contain a number of lines which display P Cygni profiles (see, e.g., 
Fig.~\ref{fig:ism}) due to the stellar wind. \citet{kf1985} determined a 
terminal wind velocity of $v_\infty = 3\,900\,\mathrm{km/sec}$, \citet[][HST/GHRS and IUE observations]{kdr1998} 
determined mass-loss rates of $\dot{M}=10^{-7.3}\,\mbox{M}_\odot/\mathrm{yr}$ from 
C\,{\sc iv} $\lambda\lambda$\,1548.2, 1550.8\,\AA\
and \citet[][ORFEUS]{kw1998} measured $\dot{M}=10^{-7.7}\,\mbox{M}_\odot/\mathrm{yr}$ from 
O\,{\sc vi} $\lambda\lambda$\,1031.9, 1037.6\,\AA. 

In order to reproduce the observations, we have to model both, the photospheric spectrum as well as the ISM 
absorption lines. For our photospheric models, we adopt the parameters by \citet[][see above]{dwk1997}.
We employed {\it TMAP}, the T\"ubingen NLTE Model Atmosphere Package \citep{wea2003}, to calculate 
those photospheric lines which are not affected by the stellar wind. 
We consider opacities of all elements from H to Ni.

For the calculation of lines which display P Cygni profiles we used the wind code by L\@. Koesterke (priv\@. comm.).
From Ne\,{\sc vii} $\lambda\,973.3\,\mathrm{\AA}$ (Fig.~\ref{fig:ism}), we determined 
$v_\infty = 3\,500\,\mathrm{km/sec}$ and $\dot{M}=10^{-7.7}\,\mbox{M}_\odot/\mathrm{yr}$.
This is in agreement with the values given by \citet{kdr1998} and \citet{kw1998}.

The ISM absorption lines are modeled with the {\sc Owens} program which takes into account different radial and turbulent
velocities, temperatures, chemical compositions, as well as column densities for each included element. 

In Fig.~\ref{fig:ism}, we show a comparison of our synthetic spectra with the observations.
The combination of photospheric and ISM models is in good agreement with the observations 
and fine-tuning of the parameters will improve the result.

Our spectral analysis of the STIS and FUSE observation of the CSPN NGC\,7094 (Fig.~\ref{fig:feni}) 
confirms the Fe underabundance found by \citet{mea2002}. We can determine an even lower upper limit of 
0.025 times the solar abundance.

\section*{Conclusions}
One explanation for the iron-deficiency and the non-existing nickel
overabundance in NGC\,7094 could be that iron was transformed by
n-captures to even heavier elements than nickel. We presently are unable
to test this idea because of the lack of atomic data.

We would like to stress that, in contrast to the other PG\,1159 stars, the
mere iron-deficiency in NGC\,7094 is particularly mysterious, because it
is a hybrid-PG\,1159 star. We recall that the hybrid-PG\,1159 stars are the
outcome of an AFTP (AGB final thermal pulse) with residual hydrogen (H$
= 0.36 $) from the envelope that was mixed with intershell matter
(Werner et al., these proceedings). One would expect to see at least the
iron that was contained in the convective H-envelope, hence 
Fe/H $\approx$\,solar, i.e., $1.6\cdot 10^{-3}$ (by mass). This is not the
case in NGC\,7094: the upper limit for Fe/H ratio is much smaller: $3.9\cdot 10^{-5}$.
\vspace{1mm}

\acknowledgements
T.R\@. is supported by the \emph{German Astrophysical Virtual Observatory} project
of the German Federal Ministry of Education and Research (BMBF) under grant 05\,AC6VTB. 
E.R\@. is supported by DFG grant We1312/30$-$1. 
J.W.K\@. is supported by the FUSE project, funded by NASA contract NAS5$-$32985.
We used the profile fitting procedure {\sc Owens} developed by M\@. Lemoine and the FUSE French Team.

\end{document}